\documentclass{ws-ijmpa}
\usepackage{epsfig,amssymb}
\begin{document}

\markboth{S.~Gureev and S.~Troitsky}
{Physical conditions in active galaxies correlated with Auger events}

\catchline{}{}{}{}{}
%
%%%%%%%%%%%%%%%%%%%%%%%%%%%%%%%%%%%%%%%%%%%%%%%%%%%%%%%%%%%%%%%%%%%%

\title{\bf PHYSICAL CONDITIONS IN
NEARBY ACTIVE GALAXIES CORRELATED WITH ULTRA-HIGH-ENERGY COSMIC RAYS
DETECTED BY THE PIERRE AUGER OBSERVATORY}

\author{
SERGEY GUREEV
}
\address{
M.V.~Lomonosov Moscow State University,\\
Moscow 119992, Russia}
\author{
SERGEY TROITSKY
}
\address{
Institute for Nuclear Research of the Russian Academy of
Sciences,\\
60th October Anniversary Prospect 7a, 117312, Moscow, Russia\\
st@ms2.inr.ac.ru
}
\date{}
\maketitle

\begin{history}
%\received{Day Month Year}
%\revised{Day Month Year}
\end{history}

\begin{abstract}
We analyze the active-galaxy
correlation reported in 2007 by the Pierre Auger Collaboration. The signal
diminishes if the correlation-function approach (counting all
``source--event'' pairs and not only ``nearest neighbours'') is used,
suggesting that the correlation may reveal individual sources and not
their population. We analyze available data on physical conditions in
these individual correlated sources and conclude that acceleration of
protons to the observed energies is hardly possible in any of these
galaxies, while heavier nuclei would be deflected by the Galactic magnetic
field thus spoiling the correlation. Our results question the Auger
interpretation of the reported anisotropy signal but do not contradict to
its explanation with intermediate-mass nuclei accelerated in Cen~A.

\keywords{Ultra-high-energy cosmic rays; active galaxies; supermassive
black holes}
\end{abstract}

%%% PACS numbers
\ccode{PACS numbers: 98.70.Sa}

%\pagestyle{myheadings}

%\tableofcontents

\section{Introduction}
\label{sec:intro}
Astronomy of ultra-high-energy (UHE; energy $E \gtrsim 10^{19}$~eV)
cosmic rays (CRs) is a new branch of science whose birth we are currently
witnessing. To identify UHECR sources is therefore an important and
challenging task. Observational evidence in favour of particular sources
is limited by low statistics, by systematic uncertainties in determination
of the primary-particle type and energy, by deflection of charged
particles in unknown cosmic magnetic fields and by other confusing effects
in the analysis. Theoretically, several compelling mechanisms of UHECR
acceleration have been developed and a number of potential astrophysical
sources have been suggested (see e.g.\ Refs.~\cite{sources,comparative}
for reviews and summary). Physical conditions in potential sources are
often uncertain thus preventing one from firm identification of
theoretically most favoured candidates.

Recently, considerable interest has been attracted by a claim of
correlation of positions of nearby active galaxies with arrival directions
of UHECR particles detected by the Pierre Auger Observatory
(PAO)~\cite{PAOagn,PAOagnLong}. The correlation was interpreted as an
evidence that UHECR particles are protons either from nearby active
galactic nuclei (AGN) or from other sources with a similar distribution in
space. The AGN interpretation was considered likely based on the previous
works favouring AGN as possible UHECR accelerators. Notably,
Ref.~\cite{PAOagnLong} cites Ref.~\cite{BiermanStr} in this context; the
latter reference however studies the most powerful AGN (BL Lac type
objects and optically violent variable quasars), while the analysis of
Ref.~\cite{PAOagn} is based on the nearby objects listed in the
V\'eron-Cetti\& V\'eron catalog~\cite{Veron} which are mostly low-power
Seyfert galaxies. In Ref.~\cite{paper1},
we analyzed general constraints from geometry of the source and
from radiation losses which restrict potential astrophysical accelerators
and compared them to the most recent astrophysical data. Our study
suggested that only the most powerful AGN are able to accelerate UHE
particles while Seyfert galaxies {\em generically} are not. In this work,
we apply the general constraints discussed in Ref.~\cite{paper1} to {\em
particular} active galaxies correlated with Auger events.

The rest of the paper is organised as follows.
In Sec.~\ref{sec:PAO:corr}, we briefly review the Auger result
emphasising particular details of the analysis of
Refs.~\cite{PAOagn,PAOagnLong}
and demonstrate that the correlation
signal is strong only if a single nearest-neighbour potential source is
counted for each event and diminishes in the correlation-function approach
when every pair ``event--source'' is counted both in the data and in the
simulations. We then select these nearest-neighbour source candidates
(mostly Seyfert galaxies) for the Auger events and study, in
Sec.~\ref{sec:PAO:individual}, their properties. We conclude that it is
unlikely that these objects can accelerate protons up to the observed
energies and discuss implications of this conclusion in
Sec.~\ref{sec:concl}.

\section{Summary and discussion of the Auger correlations}
\label{sec:PAO:corr}

\subsection{Anisotropy reported by PAO\protect\footnote{See Note Added
for discussion of the most recent results.}}
\label{sec:PAO:rev}
Recently, the Pierre Auger Collaboration reported a significant deviation
from isotropy of the arrival directions of the highest-energy cosmic rays
observed by their surface detector~\cite{PAOagn}. They studied an
(unpublished) set of events with energies $E \ge 4\times 10^{19}$~eV
and cross-correlated their arrival directions with positions of active
galaxies from the V\'eron catalog~\cite{Veron}. Three cuts were tuned to
maximize the signal: (1)~the minimal reconstructed energy of a cosmic-ray
particle, $E_{\rm min}$; (2)~the maximal redshift of a galaxy, $z_{\rm
max}$; (3)~the maximal angular separation $\psi$ between a galaxy and a
cosmic-ray arrival direction at which they are counted as correlated. An
excess of correlated event--galaxy pairs over the number expected for
isotropic distribution was found; the signal was maximized for $E_{\rm
min}=5.6 \times 10^{19}$~eV, $z_{\rm max}=0.018$ and $\psi=3.1^\circ$.

After the signal had been verified in an independent data set with the
same cuts (the probability of chance correlation in the independent set of
12 events was found to be $P\sim 1.7 \times 10^{-3}$), details of the
correlations were studied in the full data set~\cite{PAOagnLong}.
The full set was again subject to the same scans
in $E_{\rm min}$, $z_{\rm max}$ and $\psi $; the minimal formal probability
of $P_{\rm min} \sim 5\times 10^{-9}$ occured at $E_{\rm min}=5.7 \times
10^{19}$~eV, $z_{\rm max}=0.017$ and $\psi=3.2^\circ$.
The part of the data set which maximizes
the correlation signal (27 events) was published~\cite{PAOagnLong} and may
be used for analysis and further studies while the full data set remains
unavailable. To estimate $P$ from $P_{\rm min}$, one may use the so-called
penalty factor (see e.g.\ Ref.~\cite{TT:penalty}) which compensates for
the artificial signal obtained by tuning the cuts or, in other words, for
multiple tries (see also Ref.~\cite{FinleyWesterhoff} for discussion).
This penalty factor should be determined by Monte-Carlo
simulations~\cite{TT:penalty} which are however quite lengthy for these
low $P_{\rm min}$. One may estimate it roughly by counting the number of
tries: the scan in $0 \le z_{\rm max} \le 0.024$ in steps of 0.001 gives
25 tries, the scan in $1^\circ \le \psi \le 8^\circ$ in steps of
$0.1^\circ$ gives 71 tries and the scan in $E_{\rm min} \le 4 \times
10^{19}$~eV with unspecified step gives effectively the number of tries
equal to the number of events in the sample, that is 27. Altogether, the
penalty factor is roughly $25 \times 71 \times 27 \approx 4.8 \times 10^4$
(in practice, the penalty factor should be somewhat lower because the
tries are not independent) and $P_{\rm est} \sim P_{\rm min} \times 4.8
\times 10^4 \approx 2 \times 10^{-4}. $

The consistency in probabilities ($P_{\rm est} \sim P$ up to an
order-of-magnitude correction which may be attributed to
unpublished\footnote{Or published elsewhere, e.g.\ in
Ref.~\cite{PAO:BLL}.}
tries with other catalogs) and in the values of the tuned cut parameters
in the full data set and in the first half supports that a significant
deviation from isotropy, and not an outstanding chance
fluctuation, was indeed observed. The interpretation of this anisotropy is
however not unique. Firstly, active galaxies may serve only as a
tracer of the actual sources which could be distributed in the Universe in
a similar way --- indeed, most potential sources follow the distribution
of galaxies which locally is not uniform due to superclusters and voids.
This interpretation cannot be distinguished from the simplest hypothesis
that the AGN are cosmic-ray sources based on the present-level correlation
signal; both are unified in the {\em AGN hypothesis} of the Auger
collaboration: most of the cosmic rays reaching the Earth in that energy
range are protons from nearby astrophysical sources, either AGN or other
objects with a similar spatial distribution~\cite{PAOagn}.  This
interpretation requires a large (of order a hundred~\cite{PAOagnLong})
number of sources.

Secondly, it has been shown~\cite{Comment,Fargion,Wolfendale,Comment2}
that a similar signal may be easily produced in case of just a few nearby
sources but larger deflection angles (consistent with the chemical
composition of the cosmic rays observed by the fluorescent detector of the
same Pierre Auger Observatory~\cite{PAO:comp} and by the Yakutsk
experiment~\cite{Yak:2peaks}). In particular, the origin of a significant
part of cosmic-ray particles in the nearby FR~I radio galaxy Cen~A may
explain~\cite{Comment} the observed signal for the angular spread $\sim
10^\circ$ typical for intermediate-mass nuclei arriving from a source
with these Galactic coordinates~\cite{Comment2}. Moreover, this kind of
interpretation is actually supported by statistical analyses of the global
distribution of the arrival directions of the cosmic rays in the maximally
correlated Auger sample~\cite{Comment}. Qualitatively, under the AGN
hypothesis, a similar number of events is expected from the directions of
the Virgo and Centaurus superclusters (the latter is significantly farther
but the former is observed by Auger with much lower exposure). In
practice, however, no events came from Virgo while the Centaurus region
dominates the correlation~\cite{Comment}. Alternative interpretations of
the Auger signal have been considered also in
Refs.~\cite{Lemoin,Dedenko,Moskalenko,Swift-BAT,Stanev,Ghissellini,Nagar}.

Thirdly, independent experiments disagree about the presence of a similar
correlation in their data. Using a dataset of a similar size, the High
Resolution Fly's Eye experiment (HiRes) did not find a correlation with
nearby V\'eron AGN either by applying the PAO prescription or performing
similar scans in $E_{\rm min}$, $z_{\rm max}$ and $\psi$~\cite{HiRes:AGN}.
The data of the Yakutsk Extensive Air-Shower Array support the
correlation~\cite{Yakutsk:AGN}; the strongest signal occurs at $E_{\rm
min}$, $z_{\rm max}$ and $\psi$ close to the values prescribed by PAO
though the angular resolution of the experiment and possibly the energy
scale are different. We note that UHECR correlations with nearby Seyfert
galaxies from catalog~\cite{Lipovetsky} have been discovered by Uryson in
1999~\cite{Uryson,Uryson1} using the AGASA and Yakutsk data for somewhat
lower energies $E \gtrsim 4\times 10^{19}$~eV. However, a similar sample
of Seyfert galaxies from the V\'eron catalog does not correlate with the
same cosmic-ray sample~\cite{comparative}.

In this paper, we analyse the observed correlation from a completely
different point of view.

\subsection{Nearest neighbour or correlation function?}
\label{sec:only-nearest}
In any correlation analysis similar to that of~\cite{PAOagn}, the number
of observed pairs ``event--source'' is compared to that expected from
chance coincidence. The definition of the ``number of pairs'' is however
not unique: when an event is found within the angular
distance $\psi$ from $k$ sources and $k>1$, the ``number of pairs'' may be
counted either as 1 (counting only {\bf nearest-neighbour} source for each
event) or as $k$ (counting {\bf every source--event pair}). Clearly, the
same rules should be applied both to the data set and to the Monte-Carlo
sets used to estimate the expected number. While the nearest-neighbour
approach may be psychologically favoured because a single cosmic-ray
particle cannot originate in several sources, it is the second approach
which is widely applied in statistical studies by terms of the
{\bf correlation function} between two sets of points (events and
sources)\footnote{This correlation-function approach has been used in
many previous UHECR correlation studies, see e.g.\
Refs.~\cite{comparative,TT:BL,GTTT:HiRes-BL}}.

For an isotropic distribution of potential sources on the sky (when
clustering of sources at the angular scale $\sim \psi$ is consistent with
chance coincidence), both approaches give qualitatively similar results and
both may serve to test the association between the ensemble of sources and
the ensemble of cosmic rays. When sources are clustered at the angular
scale $\sim \psi$ (which is the case for the AGN sample used by PAO), it is
the correlation-function approach which should be used to study the
correlation between ensembles, while the nearest-neighbour method may
still be used to determine particular sources. If the correlation exists
with the population of (clustered) sources, one cannot say which
particular objects are responsible for individual events; the correlation
is a collective effect of many sources separated from the cosmic-ray
arrival directions by angles $\sim\psi$. A similar effect is expected if
the sample of the sources under study is just a tracer of other sources
with a similar distribution. On the other hand, if the
correlation-function approach reveals no signal but the nearest-neighbour
approach does reveal some, this is a strong argument that some of these
nearest-neighbour sources may be related to particular events (we will see
below that it is precisely what happens with the Auger correlation). This
statement, to be discussed in detail elsewhere, may be easily understood
by comparison of two cases: a large number of weak sources versus a small
number of strong sources in the sample.

\begin{figure}
\begin{center}
\includegraphics [width=0.8\textwidth]{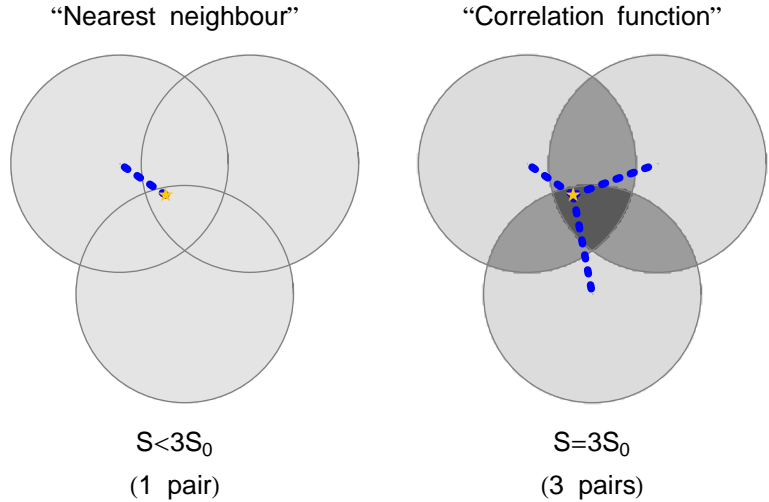}
\end{center}
\caption{
\label{fig:OnlyNearestOrNot}
Illustration of two approaches to estimate the correlation signal for
clustered sources. Suppose that the circles of radius $\psi$ (area
$S_0$) around $k>1$ potential sources (grey circles on the plot; $k=3$ in
this example) overlap and the cosmic-ray arrival direction (star) falls in
the overlapping region. In the nearest-neighbour approach (left panel),
each cosmic-ray event is associated with only one source; the area $S$
covered by circles on the sky is counted once independently of overlaps.
In the correlation-function approach (right panel), each pair
``source--event'' is counted ($k=3$ pairs in the example) and the
overlapping area is counted $k$ times in the calculation of $S$.}
\end{figure}

The key point of the calculation of the probability of chance correlation
used in PAO papers~\cite{PAOagn,PAOagnLong} is the calculation of the
fraction of the (exposure-weighted) sky area covered by circles with
radius $\psi$ centered on the potential sources in the catalog. This
fraction, called $p$ in Refs.~\cite{PAOagn,PAOagnLong}, allows to estimate
the probability of chance correlation in the nearest-neighbour approach
simply by the binomial distribution, cf.\ Ref.~\cite{PAOagn}. As it is
illustrated in Fig.~\ref{fig:OnlyNearestOrNot},
$p$ may be calculated in two different ways: in the ``nearest-neighbour''
approach, the total covered area of the sky is counted independently of
overlaps, while in the correlation-function approach, the overlap area of
two circles is counted twice.

Let us study, by means of the two methods, the sets of 27 cosmic-ray
events and of 422 AGN which maximize the Auger correlation. In the
nearest-neighbour approach and for $\psi=3.2^\circ$, we find $p\approx
0.21$, the number of correlated pairs is 20 at 5.6 expected by chance, so
$P_{\rm min}\approx 5 \times 10^{-9}$ in agreement with the Auger claim.
In the correlation-function approach, $p\approx 0.336$, the number of
correlated pairs is 25 at 8.4 expected by chance, so $P_{\rm min} \approx
1.7 \times 10^{-3}$, several orders of magnitude higher\footnote{In the
latter case, the binomial formula used in Refs.~\cite{PAOagn,PAOagnLong}
is not applicable; $P_{\rm min}$ is estimated by the Monte-Carlo
simulations.}. In the case under discussion, these two
%results
differ qualitatively: when $P_{\rm min}$ is multiplied by the penalty
factor $\sim 10^4$, the nearest-neighbour mehtod gives, as discussed
above, a significant signal, while the correlation-function result is
consistent with absence of any correlation.

This particularly strange pattern of correlation signals suggests that, if
the signal is not a chance fluctuation, then the ``nearest neighbours'' of
correlated events should be the actual sources of cosmic rays. In the rest
of the paper, we will therefore study the properties of these ``nearest
neighbours''.

\section{Physical conditions in the correlated galaxies}
\label{sec:PAO:individual}
The list of the sources responsible for the Auger correlation is given in
Table~\ref{tab:list}. In this section, we analyze available data on
physical conditions in these galaxies and compare them with requirements
for UHECR acceleration discussed in Ref.~\cite{paper1}.

\begin{table}[ph]
\tbl{
Active galaxies correlated with Auger events.
Col.~(1) gives the source name, Col.~(2) gives the energy(ies) of the
cosmic-ray particles for which this source is the ``nearest neighbour'' in
the correlated sample; Col.~(3) gives the distance to the object (taken
from NED~\protect\cite{NED}); in Cols.~(4) to (6) objects detected in
radio
(R), X rays (X) and gamma rays with $E>1$~MeV ($\gamma$) are marked with
$+$; Col.~(7) gives the AGN type (mostly from NED: Seyfert galaxies (Sy),
galaxies with H~II regions (HII), emission-line dwarf galaxy (EmG) and
FR~I radio galaxy (FRI)); Col.~(8) gives the estimated mass of the central
black hole (see Sec.~\ref{sec:PAO:individual:BH} and Table~\ref{tab:MBH}
for details and error bars). Objects with relativistic jets (J) and
non-relativistic outflows (O) are marked in Col.~(9), and the object with
knots (K), hot spots (H) and lobes (L) is marked in Col.~(10).
}
%{
{\begin{tabular}{@{}cccccccccc@{}} \toprule
Object&
$E$ &
$d$ &
\multicolumn{3}{c}{detection}&
AGN&
$\log\!\!\left(\!\frac{M_{\rm BH}}{M_\odot}\!\! \right)$ &
jets&
knots,
\\
&
(EeV)&
(Mpc)&
R&
X&
$\gamma$&
type&
& &
lobes\\
(1)&
(2)&
(3)&
(4)&
(5)&
(6)&
(7)&
(8)&
(9)&
(10)
\\
\colrule
ESO~383-G18   & 84    &51&$+$&$+$&$-$& Sy2  & 6.60 &  O & -- \\
4U~1344-60    & 66    &51&$-$&$+$&$-$& Sy1.5& 6.39 &  --& -- \\
ESO~139-G12   & 83, 59&68&$-$&$-$&$-$& Sy2  & 7.23 &  --& -- \\
IC~4518A      & 63    &66&$+$&$+$&$-$& Sy2  & 5.77 &  --& -- \\
NGC~424       & 84    &46&$+$&$+$&$-$& Sy2  & 7.34 &   O& -- \\
NGC~4945      & 58    &5 &$+$&$+$&$-$& Sy2  & 6.15 &  O & -- \\
IC~5169       & 57    &42&$+$&$-$&$-$& Sy2  & 7.52 &  O & -- \\
CGCG~374-029  & 59    &58&$+$&$-$&$-$& Sy1  & 6.01 &  --& -- \\
NGC~1346      & 85    &54&$+$&$-$&$-$& Sy1  & 7.38 &  --& -- \\
NGC~7591      & 83    &69&$+$&$-$&$-$& Sy   & 7.58 &  --& -- \\
NGC~1358      & 69    &54&$+$&$+$&$-$& Sy2  & 7.71 &  O & -- \\
Cen~A         & 69, 70&5 &$+$&$+$&$+$& FRI  & 7.65 &  J & K,H,L \\
PC~2207+0122  & 58, 71&55&$-$&$-$&$-$& EmG  &$<$5.76 &  --& -- \\
NGC~2989      & 64    &56&$+$&$-$&$-$& HII  & 6.79 &  --& -- \\
NGC~1204      & 78    &57&$+$&$-$&$-$& HII  & 7.75 &  --& -- \\
NGC~7130      & 90    &65&$+$&$+$&$-$& Sy2  & 7.35 &  O & -- \\
NGC~5506      & 83    &29&$+$&$+$&$-$& Sy2  & 6.70 &  O & -- \\
\botrule
\end{tabular}
\label{tab:list}
  }
\end{table}
\subsection{The central black holes}
\label{sec:PAO:individual:BH}
We start with the analysis of conditions close to the central black holes
of the correlated galaxies. Following the results of
Ref.~\cite{paper1}, we estimate the maximal possible energy of
UHECR particles with atomic mass $A$ and charge $Z$  from the central
black hole of mass $M_{\rm BH}$:
\begin{equation}
E_{\rm max}\simeq 3.7 \times 10^{19}~{\rm eV} \frac{A}{Z^{1/4}}
\left(  \frac{M_{\rm BH}}{10^8 M_\odot}  \right)^{3/8}.
\label{Eq:Emax-MBH}
\end{equation}

For a few objects, the mass of the black hole has been estimated
previously in the literature; for others, we
estimate $M_{\rm BH}$ from available
data.

Methods of determination of the mass $M_{\rm BH}$ of the AGN central black
hole have been reviewed in Ref.~\cite{FF} (for a recent update, see
e.g.~\cite{0712.1630}), where basic corresponding references can be found.
For 3 of 17 galaxies correlated with Auger events, $M_{\rm BH}$ has been
estimated in the literature using observations of circumnuclear dynamics
by means of:

(GD): resolved {\bf gas dynamics};

(MM): observations of accretion-disk {\bf megamasers};

(H$\alpha $): width of the {\bf H$\alpha $ line}.

For other galaxies in the sample we used, depending on the data
availability, the following methods.

(SVD): {\bf stellar velocity dispersion}. This quantity, $\sigma $, was
found to be correlated with $M_{\rm BH}$; we use the
relation~\cite{0712.1630}
$$
\log \left(\frac{M_{\rm BH}}{10^8 M_\odot}    \right)=
\left( -0.08 \pm 0.02   \right)
+
\left( 3.93 \pm 0.10   \right)
\log \left(\frac{\sigma }{200~{\rm km}\,{\rm s}^{-1}}  \right) .
$$
For some of the galaxies, $\sigma $ is given in the LEDA
database~\cite{LEDA}; for a few others, we found it in the literature.

(KSB): {\bf buldge $K_s$ magnitude}. It is also correlated with $M_{\rm
BH}$; we use the relation~\cite{0510694}
$$
\log \left(\frac{M_{\rm BH}}{10^8 M_\odot}    \right)=
\left( -2.5 \pm 0.6   \right)
+
\left(-0.45 \pm 0.03   \right) K_s^0,
$$
where the absolute bulge magnitude $K_s^0$ is related to the observed
bulge magnitude $K_s^B$ through
$$
K_s^0=K_s^B+A_K+{\rm DM}   .
$$
Here $A_K$ is the correction for the Galactic extinction in the K
band~\cite{GalacticExtinction} and DM is the distance modulus (both $A_K$
and DM are given in NED). For some of the galaxies, $K_s^B$ was determined
by the disk-bulge decomposition in previous studies; when known, it is
often quoted in NED.

(KSM): {\bf bulge $K_s$ from morphology}. When $K_s^B$ is not directly
available, one may estimate it from the total $K_s$ magnitude and the
Hubble morphological type $T$ of a galaxy by using the following
relation~\cite{0510694},
$$
K_s^B=K_s+0.297
\left( T+5 \right)
-0.040
\left( T+5 \right)^2
+0.0035
\left( T+5 \right)^3.
$$
The precision of this relation is of order $0.5^{\rm m}$ and depends on
$T$~\cite{0510694}. While this method is the least precise, it may be used
to estimate $M_{\rm BH}$ for any galaxy with known distance because either
$K_s$ or a lower limit on it may be determined from the 2MASS full-sky
survey data. For 9 of 17 correlated galaxies, KSM is the only available
method to estimate $M_{\rm BH}$.

\begin{figure}
\begin{center}
\includegraphics [width=0.35\textwidth]{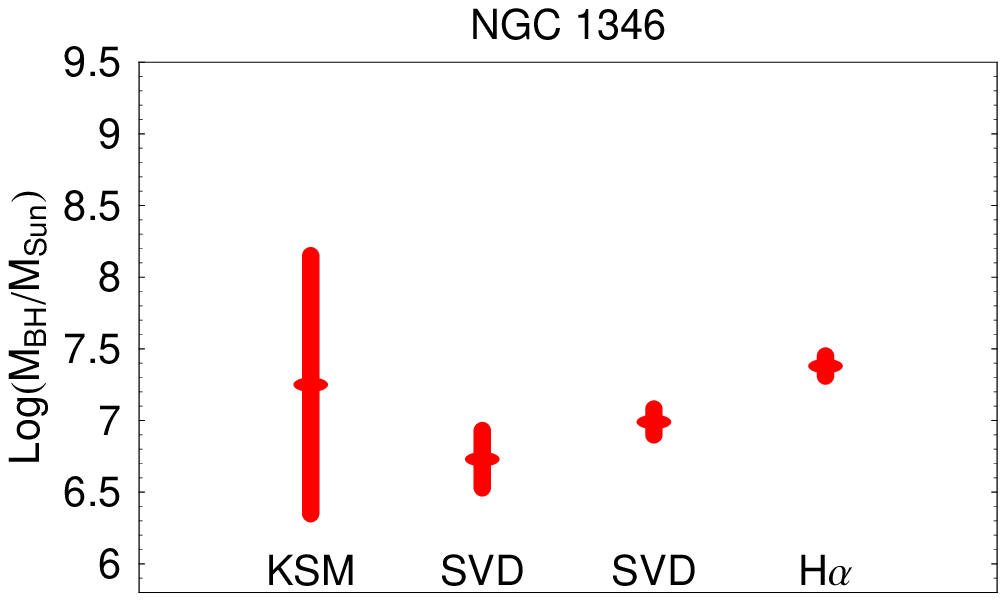}%
\includegraphics [width=0.35\textwidth]{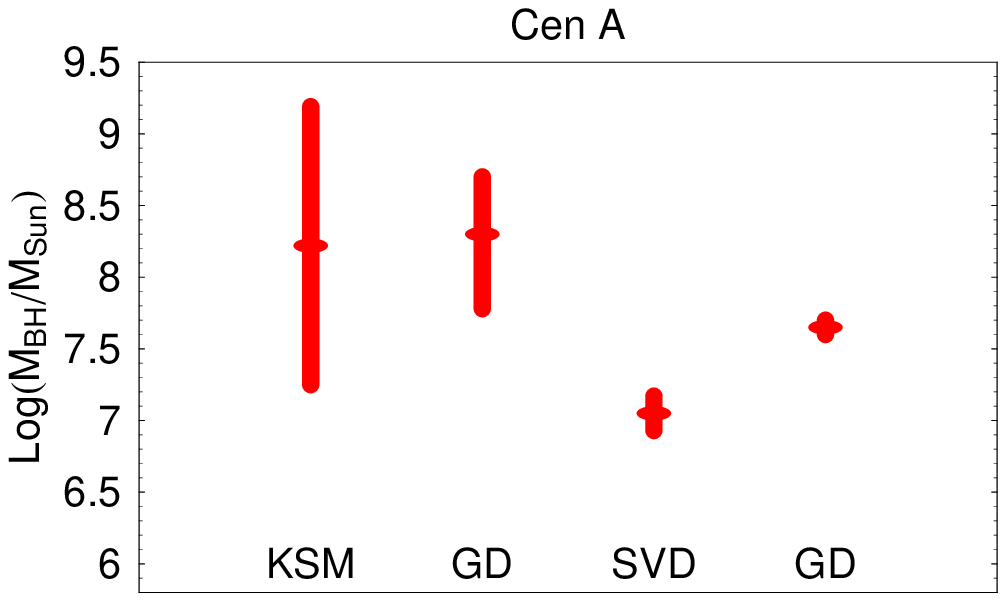}
\end{center}
\caption{
\label{fig:M_BHestimates}
Different estimates of the mass of the central black hole in two galaxies
studied in Sec.~\ref{sec:PAO:individual}. Letters denote the method of
estimation (see Sec.~\ref{sec:PAO:individual:BH}).}
\end{figure}

The data we use are collected in Table~\ref{tab:MBH-dat}
\begin{table}
\tbl{
Data used to estimate the mass of the central black hole in active
galaxies correlated with Auger events. Col.~(1): the source name.
Col.~(2): $K_s$ magnitude (A: $95\%$ confidence-level lower limit from the
2MASS point source rejection table; B: bulge decomposed magnitude from
Ref.~\protect\cite{Kband}, C: buldge decomposed magnitude from
Ref.~\protect\cite{0510694}, D: DENIS value from LEDA, no letter: 2MASS
value from NED; for B and C, 2MASS error bars assumed). Col.~(3):
correction for Galactic extinction in $K$ band
(Ref.~\protect\cite{GalacticExtinction}, NED). Col.~(4): distance modulus
(NED; accuracy is $\pm 0.15^{\rm m}$ except for PC~2207+0122 ($\pm
0.21^{\rm m}$). Col.~(5): Hubble morphological type from LEDA. Col.~(6):
central velocity dispersion (star denotes data from
Ref.~\protect\cite{Kband} for which 10\% accuracy was assumed in
calculations, if no reference given, data from LEDA). }
{
\begin{tabular}{@{}cccccc@{}}
\toprule
Object&
$K_s$, mag&
$A_K$, mag&
DM, mag&
$T$ &
$\sigma$, km/s \\
(1)&
(2)&
(3)&
(4)&
(5)&
(6)
\\
\colrule
ESO~383-G18   & $  11.99 \pm  0.05 $ B& 0.022&  33.52 & 0 & 92.3 *  \\
4U~1344-60    & $  11.140\pm  0.043$  & 1.082&  33.49 & 5 & --      \\
ESO~139-G12   & $  9.709 \pm  0.039$  & 0.027&  34.17 & 4 & --      \\
IC~4518A      & $  11.09 \pm  0.22 $ D& 0.058&  34.10 & 6 & --      \\
NGC~424       & $  11.23 \pm  0.02 $ B& 0.006&  33.31 & 0 & 142.6 * \\
NGC~4945      & $  4.483 \pm  0.017$  & 0.065&  28.43 & 4 & $127.9\pm 19.1$\\
IC~5169       & $  9.776 \pm  0.029$  & 0.006&  33.10 & 0 & --           \\
CGCG~374-029  & $  11.260\pm  0.050$  & 0.043&  33.80 & 5 & --           \\
NGC~1346      & $  9.790 \pm  0.029$  & 0.019&  33.66 & 3 & $116.1\pm
4.7$~\cite{0712.1630}
 \\
NGC~7591      & $  9.602 \pm  0.032$  & 0.038&  34.19 & 3 & --           \\
NGC~1358      & $  10.18 \pm  0.003$ C& 0.023&  33.65 & 0 & $176.7\pm 10.1$\\
Cen~A         & $  3.942 \pm  0.016$  & 0.042&  28.39 & -2&  $119.8\pm 7.1$\\
PC~2207+0122  & $ >17.057          $ A& 0.013&  33.69 & ? & --          \\
NGC~2989      & $  10.229\pm  0.049$  & 0.023&  33.72 & 4 & --          \\
NGC~1204      & $  9.947 \pm  0.030$  & 0.027&  33.77 & 0 & --          \\
NGC~7130      & $  10.18 \pm  0.01 $ B& 0.011&  34.08 & 1 & 143.2 *     \\
NGC~5506      & $  10.74 \pm  0.015$ B& 0.022&  32.29 & 1 & 97.9 *      \\
\botrule
\end{tabular}
\label{tab:MBH-dat}
}
\end{table}
while the summary
of $M_{\rm BH}$ estimates is given in Table~\ref{tab:MBH}.
\begin{table}
\tbl{
The estimates of the mass of the central black hole of active galaxies
correlated with Auger events. Col.~(1) gives the source name, Col.~(2)
gives the mass estimate, Col.~(3) gives the method (see
Sec.~\protect\ref{sec:PAO:individual:BH}), Col.~4 gives the reference to
either the mass value (v) or the data used to calculate it (d). If no
reference is given, the LEDA value of $\sigma$ was used in the SVD method,
and the data of Table~\protect\ref{tab:MBH-dat} were used in the KSB, KSM
method. The most precise estimates for each object, which we use in this
work, are given in the bold face.} {
\begin{tabular}{@{}cccc@{}}
\toprule
Object&
$\log\left(\frac{M_{\rm BH}}{10^8 M_\odot} \right)$ &
Method&
Reference
\\
(1)&
(2)&
(3)&
(4)
\\
\colrule
ESO~383-G18 & $\bf -1.40 \pm 0.19$ & SVD & \cite{Kband} d
 \\
            & $-0.80 \pm 0.88$ & KSB &                  \\
4U~1344-60  & $\bf -1.61 \pm 0.92$ & KSM &                  \\
ESO~139-G12 & $\bf -0.77 \pm 0.94$ & KSM &                  \\
IC~4518A    & $\bf -2.23 \pm 0.86$ & KSM &                  \\
NGC~424     & $\bf -0.66 \pm 0.18$ & SVD & \cite{Kband} d
 \\
            & $-0.56 \pm 0.90$ & KSB &                  \\
NGC~4945    & $\bf -1.85 \pm 0.18$ & MM  & \cite{ApJ_481_L23} v
 \\
            & $-0.98 \pm 0.93$ & KSM &                  \\
            & $-0.84 \pm 0.26$ & SVD &                  \\
            & $<-2           $ & GD  & \cite{0709.3960} v
 \\
IC~5169     & $\bf -0.48 \pm 0.92$ & KSM &                   \\
CGCG~374-029& $\bf -1.99 \pm 0.91$ & KSM &                   \\
NGC~1346    & $\bf -0.62 \pm 0.07$ & H$\alpha$ & \cite{0712.1630} v
 \\
            & $-0.75 \pm 0.90$ & KSM &                   \\
            & $-1.27 \pm 0.20$ & SVD &                   \\
            & $-1.01 \pm 0.09$ & SVD & \cite{0712.1630} d
 \\
NGC~7591    & $\bf -0.42 \pm 0.92$ & KSM &                   \\
NGC~1358    & $\bf -0.29 \pm 0.10$ & SVD &                   \\
            & $ 0.14 \pm 0.95$ & KSM &                   \\
            & $ 0.07 \pm 0.93$ & KSB & \cite{0510694} d
 \\
Cen~A       & $\bf -0.35 \pm 0.05$ & GD  & \cite{0709.1877}  v
 \\
            & $ 0.22 \pm 0.97$ & KSM &                    \\
            & $-0.95 \pm 0.12$ & SVD &                    \\
            & $0.30^{+0.40}_{-0.52}$ & GD  & \cite{ApJ_549_915} v
 \\
PC~2207$+$0122& $\bf <-2.22        $ & KSM &                    \\
NGC~2989    & $\bf -1.21 \pm 0.92$ & KSM &                    \\
NGC~1204    & $\bf -0.25 \pm 0.93$ & KSM &                    \\
NGC~7130    & $\bf -0.65 \pm 0.18$ & SVD & \cite{Kband} d
 \\
            & $ 0.26 \pm 0.93$ & KSB &                    \\
NGC~5506    & $\bf -1.30 \pm 0.19$ & SVD & \cite{Kband} d
 \\
            & $-0.79 \pm 0.89$ & KSB &                    \\
            & $-0.25 \pm 0.20$ & SVD &                    \\
\botrule
\end{tabular}
\label{tab:MBH}
}
\end{table}
For estimates of the maximal
energy of cosmic-ray particles we always use the most precise estimate of
the black-hole mass given in the bold face in Table~\ref{tab:MBH}. When
different methods are applicable, less precise estimates are in a good
agreement with better ones (a comparison plot is given in
Fig.~\ref{fig:M_BHestimates}).

\begin{figure}
\begin{center}
\includegraphics [width=0.7\textwidth]{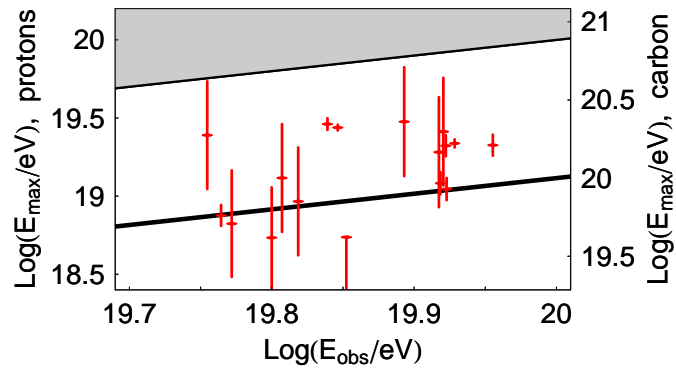}
\end{center}
\caption{
\label{fig:maxE-obsE(BH)}
The maximal energy $E_{\rm max}$ attainable by protons and carbon nuclei in
the vicinity of the central black hole of individual active galaxies
correlated with the Auger events, versus the energy $E_{\rm obs}$ of the
observed cosmic-ray particle associated with the same galaxy (if a source
corresponds to two events, $E_{\rm obs}$ is the highest one). The shaded
area corresponds to the allowed region $E_{\rm obs} \le E_{\rm max}$ for
protons; the area above the thick black line corresponds to the allowed
region $E_{\rm obs} \le E_{\rm max}$ for carbon nuclei. The error bars
correspond to precision of determination of the black-hole mass; Znajeck
conservative upper limit for the magnetic field is assumed (see
Ref.~\protect\cite{paper1} for details). For PC~2207$+$0122, a dwarf galaxy
associated with two cosmic-ray events but not detected in the 2MASS
survey, only an upper limit on $E_{\rm max}$ is available.}
\end{figure}

Figure~\ref{fig:maxE-obsE(BH)}
compares the maximal attainable energy
$E_{\rm max}$ for protons and nuclei near the central black hole of the
galaxies with actual energies of the correlated cosmic-ray events. The
energy $E_{\rm max}$ has been calculated using Eq.~(\ref{Eq:Emax-MBH}). We
remind that this equation gives an upper limit on $E_{\rm max}$ which is
lower in particular models and for realistic values of the magnetic
fields. Central black holes of any of these galaxies cannot accelerate
protons to the observed energies. While they could be able to accelerate
heavier nuclei, deflections of these nuclei in the Galactic magnetic field
would spoil the directional correlation at $\psi \sim 3^\circ$.

\subsection{Jets and other extended structures}
\label{sec:PAO:individual:jets}
In eight of the correlated galaxies, extended nuclear outflows have been
detected. Except for Cen~A, which is discussed separately in the following
section, these outflows are most probably non-relativistic, and when the
origin of the emission may be determined, it is thermal.
Observations of outflows in correlated Seyfert galaxies are summarized in
Table~\ref{tab:jets-dat}.
\begin{table}
\tbl{
Observations of outflows detected in correlated galaxies.
Col.~(1): the source name. Col.~(2): linear size.
Col.~(3): frequency of observations.
Col.~(4): observed flux.
Col.~(5): estimate of the magnetic field (for kpc wind in NGC~5506, from
thermal source model of Ref.~\protect\cite{9711137}; for other cases, from
Eq.~(\protect\ref{Eq:B-outlow})). Col.~(6): reference to the data. }
{
\begin{tabular}{@{}cccccc@{}}
\toprule
Object&
$R$, kpc&
$\nu $, GHz&
$F$, mJy&
$B_{\rm est}$, G &
Ref.\\
(1)&
(2)&
(3)&
(4)&
(5)&
(6)
\\
\colrule
ESO~383-G18   & 0.107 & 8.3 & 1.7 &$2\times 10^{-7}$&\cite{ApJ-537-152}
 \\
NGC~424       & 0.33  & 8.3 &13.1 &$ 10^{-7}$&\cite{ApJ-537-152}
 \\
              & 0.5   & 1.5 &23.9 &$5\times 10^{-8}$&\cite{9901236}
 \\
              &       & 8.4 &13.1 &$9\times 10^{-8}$&\cite{9901236}
 \\
NGC~4945      & 31.4  & 4.8 &51   &$9\times 10^{-10}$&\cite{AA-216-39}
 \\
              &0.5&$1.5\times 10^{18}$&$2.6\times 10^{-6}$&
                                       $ 10^{-7}$&\cite{0204361}
 \\
IC~5169       & 0.35  & 1.5 &17.6 &$5\times 10^{-8}$&\cite{9901236}
 \\
              &       & 8.4 & 3.7 &$6\times 10^{-8}$&\cite{9901236}
 \\
NGC~1358      & 0.36  & 1.5 & 3.4 &$3\times 10^{-8}$&\cite{9901236}
 \\
              &       & 8.4 & 0.9 &$3\times 10^{-8}$&\cite{9901236}
 \\
Cen~A         &\multicolumn{5}{c}{see Sec.~\ref{sec:PAO:individual:CenA}}\\
NGC~7130      &0.188  & 8.4 & 18.1 &$4\times 10^{-7}$&\cite{0001459}
 \\
NGC~5506      &0.0015 & 8.4 & 24.4 &$3\times 10^{-5}$&\cite{AA-417-925}
 \\
              &0.36   & 8.4 & 67.6 &$10^{-7}$&\cite{0001459}
 \\
              &$1.2\div 3.0$&$1.5\times 10^{18}$&
$(0.6\div 2.4)\!\times\! 10^{-7}$& $(0.8\div 6)\!\times\!
10^{-6}$&\cite{9711137}
 \\
              &0.302  & 8.3 &95.0  &$2\times 10^{-7}$&\cite{ApJ-537-152}
 \\
\botrule
\end{tabular}
\label{tab:jets-dat}
}
\end{table}
Magnetic fields in these outflows have not been studied; to get
order-of-magnitude estimates, we assume that the energy of the magnetic
field is the same as the total energy of the radiation field.
We assume a
spherical source of the radius $R/2$ filled with magnetic field $B$, so
that the energy of the magnetic field is $E_1\sim \pi R^3 B^2/6$. The
energy of the radiation field may be estimated as $E_2\sim R\, dE_2/dt$,
where the energy loss $dE_2/dt=4\pi D^2 \int\!I_\nu\,d\nu$ for a source at
the distance $D$ with the observed flux $I_\nu$ at the frequency $\nu$ (we
use the definition of $I_\nu$ as the energy per unit area per unit time
per unit frequency interval per unit solid angle, that is $I_\nu$ is
measured e.g.\ in Jansky). Within our precision, $\int\!I_\nu\,d\nu\sim
I_\nu \nu$. Equating $E_1=E_2$ and expressing $D/R$ through the angular
size of the source $\theta$, we arrive at
\begin{equation}
B\sim 6.5\times 10^{-6}~{\rm G}~
\left(\frac{\theta}{\rm mas} \right)^{-1}
\left(\frac{I}{\rm mJy} \right)^{1/2}
\left(\frac{\nu}{\rm GHz} \right)^{1/2}.
\label{Eq:B-outlow}
\end{equation}
Estimated magnetic fields (given in Table~\ref{tab:jets-dat}) do not
exceed those expected for normal galaxies of the same size; the outflows
certainly cannot serve as acceleration sites of UHECRs because they do
not satisfy the Hillas criterion.

\subsection{The case of {\em Centaurus~A}}
\label{sec:PAO:individual:CenA}
One of the correlated galaxies, NGC~5128, hosts a radio galaxy known as
Cen~A. Due to its proximity, it represents a textbook example of FR~I
radio galaxy, studied in great details for decades. It has been suggested
long ago as a
source of some~\cite{Cavallo,Ancho:Cen} or even most
\cite{Farrar,Weiler} of the
observed cosmic rays of extreme energies.
Both the original~\cite{PAOagn} and some of the
alternative~\cite{Comment,Fargion,Wolfendale,Comment2,Moskalenko,Nagar}
interpretations of the Auger correlation signal imply that Cen~A is a
UHECR source. We therefore devote a separate section to the study of
physical conditions in Cen~A.

Centaurus~A (for reviews, see Refs.~\cite{AA-rev-8-237,ChinJAAp-6-106} and
a continuously updated webpage~\cite{Steinler-webpage}) is a low-power
radio galaxy (its luminosity is intermediate between Seyfert and
FR~I~\cite{0707.0177}) which, most probably, experienced a major merging
event recently. High-resolution studies determine the compact nucleus
(optically classified as Seyfert~2) whose multifrequency spectral energy
distribution is very similar to those of the BL~Lac type objects (see
Refs.~\cite{0707.0177,MNRAS-256-1,0105159,ApJ-549-L173} for
discussion\footnote{Cen~A is the only object in the AGN sample studied by
PAO which is classified as a BL Lac in the V\'eron
catalog~\cite{Veron}.}); the jet which experience several bendings
observed at scales from light months~\cite{PASJ-52-102} to dozens of
kiloparsecs from the nuclei; inner radio lobes associated with the jet and
giant outer radio lobes which span about $9^\circ$ on the sky. The jet is
mildly relativistic, $\Gamma<2.5$ at parsec scales~\cite{0707.0177}
(velocity $\sim 0.45c$ was inferred~\cite{AJ-115-960} and proper motions
$\sim 0.12c$ were observed~\cite{AJ-122-1697}). Polarization studies
reveal that the magnetic field in the jet is parallel to the jet axis at
the scales of at least 3~kpc (see e.g.\ Ref.~\cite{ApJ-593-169} and
references therein). Several dosen knots are resolved in the jet; their
radio and X-ray positions are systematically offset from each other (see
e.g.\ Ref.~\cite{ApJ-569-54}). These knots are possible sites of particle
acceleration; however recent X-ray observations (e.g.~\cite{0710.1277})
suggest that another, distributed along the jet, acceleration mechanism
should be at work. This mechanism may be associated with a hard-spectrum
shear layer possibly observed along the
kiloparsec-scale jet~\cite{0510661}. Multifrequency observations (e.g.\
Ref.~\cite{0601421}) establish the synchrotron origin of the jet emission
with high confidence.

\begin{figure}
\begin{center}
\includegraphics [width=0.7\textwidth]{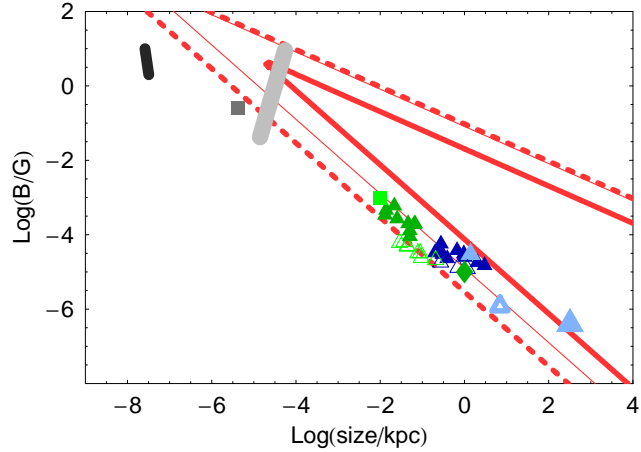}
\end{center}
\caption{
The size-field plot for different parts of Cen~A. Grey colours: the nuclear
region (light-grey error-bar diagonal, radio-to-X-ray modelling of
Ref.~\protect\cite{ApJ-238-539}; dark-grey error-bar diagonal, that of
Ref.~\protect\cite{0710.2847}; grey box, synchrotron self-absorption
measurement
of Ref.~\protect\cite{0707.0177}). Green colours: the jet (small green
triangles, minimal-pressure estimates for resolved
components~\protect\cite{ApJ-395-444}; light-green empty triangles,
equipartition
estimates for resolved components~\protect\cite{ApJ-273-128}; green box,
model estimate for a knot~\protect\cite{0709.2210}; green diamond, model
estimate for the possible hard-spectrum shear
layer~\protect\cite{0510661}).
Blue colours: the lobes (large light-blue triangle, equipartition
estimate for the giant outer lobes~\protect\cite{AA-355-863} (see also
Ref.~\protect\cite{new-lobes}); other points are estimates for inner lobes:
small light-blue triangle, minimal-energy~\protect\cite{ApJ-646-L41};
small empty light-blue triangle, equipartition~\protect\cite{AA-355-863};
dark-blue triangles, minimal-pressure~\protect\cite{ApJ-395-444};
dark-blue empty triangles, equipartition~\protect\cite{ApJ-273-128}). The
allowed region for acceleration of $7\times 10^{19}$~eV particles is
located between red lines (thick, protons; thin, carbon nuclei; dashed,
iron nuclei; lower lines corresponds to the Hillas limit, upper ones
corresponds to the radiation-loss limit for diffusive acceleration).  Note
that the diffusive-acceleration loss limit is determined by $A/Z$ and is
therefore indistinguishable for iron ($A=56$, $Z=26$) and carbon ($A=12$,
$Z=6$). See Ref.~\protect\cite{paper1} for details of the calculation of
allowed areas and for description of the methods to estimate the magnetic
fields.
\label{fig:cenA}
}
\end{figure}

Numerous estimates of the magnetic field in different parts of Cen~A are
summarized in Fig.~\ref{fig:cenA}.
Acceleration of protons to the energy of $7\times 10^{19}$~eV
(the energy of the events associated with the Cen~A nucleus by PAO) is
hardly possible in this source, except maybe for the giant outer lobes
which are however displaced from the nucleus by an angular distance
exceeding $\psi=3.2^\circ$. On the other hand, acceleration of
intermediate-mass and heavy nuclei in the jet and lobes of Cen~A remains a
plausible interpretation of the correlation signal.

\section{Conclusions}
\label{sec:concl}
The correlations observed by PAO may be explained
either by the AGN hypothesis or by chance coincidence. The AGN hypothesis
implies that the primary cosmic-ray particles are protons and has three
options:

(i) the correlated AGN are (some of) the sources;

(ii) the AGN as a whole are sources; particular
correlated AGN may not be the actual sources if they are located in
regions of enhanced AGN density;

(iii) the sources are not AGN but follow a similar distribution in space,
that is, like AGN, they follow the local large-scale structure of the
Universe.

The options (iii) and (ii) are disfavoured by the absence of the
events from the direction of the Virgo supercluster~\cite{Comment};
the options (i) and (ii) are disfavoured by the luminosity
argument \cite{Dedenko}.

In this paper we have noted that (ii) and (iii) are
also disfavoured by the disappearence of correlations in terms of the
correlation-function approach, contrary to the nearest-neighbour
probability estimate.
In order to test the option (i),
we have studied physical conditions in particular correlated
galaxies. Most of them are low-power Seyfert galaxies; we have
demonstrated that {\bf none of the correlated AGN can accelerate protons to
the observed energies}, thus excluding the option (i) and giving further
support to the explanation of correlations by means different than the AGN
hypothesis.

Our results suggest that the correlated galaxies,
notably Cen A, can accelerate nuclei to the observed energies. The
deflections would however spoil the small-angle correlations, but the
origin of a significant part of the detected particles in Cen~A jet
and lobes remains one of the most probable explanations for the observed
anisotropy.

\section*{Note added}
Recently, a few months after this paper was completed and posted in arXiv,
PAO released (see e.g.\ Ref.~\cite{new-PAO-talk}) information about tests
of the AGN correlations with new data. In agreement with our results,
these data speak against the AGN interpretation of the observed
anisotropy: among 31 events with $E>E_{\rm min}$ recorded after
publication of Ref.~\cite{PAOagn} and before March 2009, \ 8 correlate
with
AGN while 6.5 chance coincidences are expected for isotropic distribution
of arrival directions~\cite{new-PAO-talk}.

\section*{Acknowledgments}
The authors are indebted to D.~Gorbunov, G.~Rubtsov, L.~Stawarz,
P.~Tinyakov and I.~Tkachev for discussions. We acknowledge the use of
online tools~\cite{NED,LEDA,Steinler-webpage}. This work  was
  supported
in part
by the grants RFBR 07-02-00820, RFBR 09-07-00388,
NS-1616.2008.2 (ST), by FASI under state contracts 02.740.11.0244 (ST)
and 02.740.11.5092 (SG and ST) and by the Dynasty Foundation (ST).


\begin{thebibliography}{76}

\bibitem{sources}
D.~F.~Torres and L.~A.~Anchordoqui,
  %``Astrophysical origins of ultrahigh energy cosmic rays,''
  Rept.\ Prog.\ Phys.\  {\bf 67}, 1663 (2004)
  [arXiv:astro-ph/0402371].
  %%CITATION = RPPHA,67,1663;%%
\bibitem{comparative}
 D.~S.~Gorbunov and S.~V.~Troitsky,
%   ``A comparative study of correlations between arrival directions of
%   ultra-high-energy cosmic rays and positions of their potential
%astrophysical
%sources,''
  Astropart.\ Phys.\  {\bf 23}, 175 (2005)
  [arXiv:astro-ph/0410741].
  %%CITATION = APHYE,23,175;%%
\bibitem{PAOagn}
  J.~Abraham {\it et al.}  [Pierre Auger Collaboration],
%   ``Correlation of the highest energy cosmic rays with nearby
%extragalactic
%objects,''
  Science {\bf 318}, 938 (2007)
  [arXiv:0711.2256 [astro-ph]].
  %%CITATION = SCIEA,318,938;%%
\bibitem{PAOagnLong}
 J.~Abraham {\it et al.}  [Pierre Auger Collaboration],
%   ``Correlation of the highest-energy cosmic rays with the positions of
%nearby
%active galactic nuclei,''
  Astropart.\ Phys.\  {\bf 29}, 188 (2008)
  [Erratum-ibid.\  {\bf 30}, 45 (2008)]
  [arXiv:0712.2843 [astro-ph]].
  %%CITATION = ARXIV:0712.2843;%%
\bibitem{BiermanStr}
P.L.~Biermann and P.~A.~Strittmatter,
  %{\it Synchrotron emission from shock waves in active galactic nuclei,}
 Astrophys.\ J.\  {\bf 322}, 643  (1987)
  %%CITATION = ASJOA,322,643;%%
\bibitem{Veron}
M.P.~V\'eron-Cetty and P.~V\'eron,
%{\it  A catalogue of quasars and active nuclei: 12th edition,}
{Astron.\ Astrophys.}\ {\bf 455} 773 (2006)
\bibitem{paper1}
R.~Ptitsyna  and S.~Troitsky ,
%{\it
%Physical conditions in potential sources of ultra-high-energy cosmic rays.
%I.
%Updated Hillas plot and radiation-loss constraints,} 2008,
arXiv:0808.0481[astro-ph]
\bibitem{TT:penalty}
 P.~Tinyakov and I.~Tkachev,
 %  ``Cuts and penalties: comment on ``The clustering of ultra-high energy
%cosmic
%rays and their sources'',''
  Phys.\ Rev.\  D {\bf 69}, 128301 (2004)
  [arXiv:astro-ph/0301336].
  %%CITATION = PHRVA,D69,128301;%%
\bibitem{FinleyWesterhoff}
 C.~B.~Finley and S.~Westerhoff,
%   ``On the Evidence for Clustering in the Arrival Directions of AGASA's
  %Ultrahigh Energy Cosmic Rays,''
  Astropart.\ Phys.\  {\bf 21}, 359 (2004)
  [arXiv:astro-ph/0309159].
  %%CITATION = APHYE,21,359;%%
\bibitem{PAO:BLL}
 D.~Harari  [The Pierre Auger Collaboration],
 %  ``Search for correlation of UHECRs and BL Lacs in Pierre Auger
%Observatory
%data,''
  arXiv:0706.1715 [astro-ph].
  %%CITATION = ARXIV:0706.1715;%%
\bibitem{Comment}
D.~Gorbunov, P.~Tinyakov, I.~Tkachev and S.~V.~Troitsky,
%   ``Comment on 'Correlation of the Highest-Energy Cosmic Rays with Nearby
  %Extragalactic Objects',''
  JETP Lett.\  {\bf 87}, 461 (2008)
  [arXiv:0711.4060 [astro-ph]].
%%CITATION = ARXIV:0711.4060;
\bibitem{Fargion}
  D.~Fargion,
  %``Light Nuclei solving Auger puzzles ?,''
  Phys.\ Scripta {\bf 78}, 045901 (2008)
  [arXiv:0801.0227 [astro-ph]].
 %%CITATION = ARXIV:0801.0227;%%
\bibitem{Wolfendale}
T.~Wibig and A.~W.~Wolfendale,
  %``Heavy Cosmic Ray Nuclei from Extragalactic Sources above 'The Ankle',''
  arXiv:0712.3403 [astro-ph].
  %%CITATION = ARXIV:0712.3403;%%
\bibitem{Comment2}
  D.~S.~Gorbunov, P.~G.~Tinyakov, I.~I.~Tkachev and S.~V.~Troitsky,
%   ``On the interpretation of the cosmic-ray anisotropy at ultra-high
  %energies,''
  arXiv:0804.1088 [astro-ph].
  %%CITATION = ARXIV:0804.1088;%%
\bibitem{PAO:comp}
 M.~Unger  [The Pierre Auger Collaboration],
%   ``Study of the Cosmic Ray Composition above 0.4 EeV using the
%Longitudinal
%Profiles of Showers observed at the Pierre Auger Observatory,''
  arXiv:0706.1495 [astro-ph].
  %%CITATION = ARXIV:0706.1495;%%
\bibitem{Yak:2peaks}
A.~V.~Glushkov {\it et al.},
  %, I.~T.~Makarov, M.~I.~Pravdin, I.~E.~Sleptsov, D.~S.~Gorbunov,
  %G.~I.~Rubtsov and S.~V.~Troitsky,
% {\it Muon content of ultra-high-energy air showers: Yakutsk data versus
%  simulations,} 2008
JETP Lett.\  {\bf 87}, 190 (2008)
  [arXiv:0710.5508 [astro-ph]]
  %%CITATION = JTPLA,87,190;%%
\bibitem{Lemoin}
K.~Kotera and M.~Lemoine,
%   ``The optical depth of the Universe for ultra-high energy cosmic ray
  %scattering in the magnetized large scale structure,''
  Phys.\ Rev.\  D {\bf 77}, 123003 (2008)
  [arXiv:0801.1450 [astro-ph]].
  %%CITATION = PHRVA,D77,123003;%%
\bibitem{Dedenko}
L.G.~Dedenko  {\it et al.},
  %, D.~A.~Podgrudkov, T.~M.~Roganova and G.~F.~Fedorova,
%``The cosmic ray luminosity of the nearby active galactic nuclei,''
  arXiv:0804.4582 [astro-ph].
  %%CITATION = ARXIV:0804.4582;%%
\bibitem{Moskalenko}
I.~V.~Moskalenko, L.~Stawarz, T.~A.~Porter and C.~C.~Cheung,
%   ``On the Possible Association of Ultra High Energy Cosmic Rays with
%Nearby
%Active Galaxies,''
  Astrophys.\ J.\  {\bf 693}, 1261 (2009)
  [arXiv:0805.1260 [astro-ph]].
  %%CITATION = ARXIV:0805.1260;%%
\bibitem{Swift-BAT}
M.~R.~George {\it et al.},
  %A.~C.~Fabian, W.~H.~Baumgartner, R.~F.~Mushotzky and J.~Tueller,
%``On Active Galactic Nuclei as Sources of Ultra-High Energy Cosmic Rays,''
  arXiv:0805.2053 [astro-ph].
  %%CITATION = ARXIV:0805.2053;%%
\bibitem{Stanev}
T.~Stanev,
  %``A comment on the Auger events correlation with AGN,''
  arXiv:0805.1746 [astro-ph].
  %%CITATION = ARXIV:0805.1746;%%
\bibitem{Ghissellini}
G.~Ghisellini  {\it et al.},
  %, G.~Ghirlanda, F.~Tavecchio, F.~Fraternali and
  %G.~Pareschi,
 % {\it Ultra-High Energy Cosmic Rays, Spiral galaxies and Magnetars,}
   arXiv:0806.2393 [astro-ph]
%%CITATION = ARXIV:0806.2393;%%
\bibitem{Nagar}
N.~M.~Nagar and J.~Matulich,
%   ``Ultra-High Energy Cosmic Rays Detected by the Pierre Auger
%  Observatory: First Direct Evidence, and its Implications, that a Subset
%   Originate in
  %Nearby Radiogalaxies,''
  arXiv:0806.3220 [astro-ph].
  %%CITATION = ARXIV:0806.3220;%%
\bibitem{HiRes:AGN}
 R.~U.~Abbasi {\it et al.} [HiRes Collaboration],
%   ``Search for Correlations between HiRes Stereo Events and Active
%   Galactic
   %Nuclei,''
  Astropart.\ Phys.\  {\bf 30}, 175 (2008)
  [arXiv:0804.0382 [astro-ph]].
  %%CITATION = ARXIV:0804.0382;%%
\bibitem{Yakutsk:AGN}
A.~A.~Ivanov,
  %``A search for Extragalactic Sources of Ultrahigh-Energy Cosmic Rays,''
  Pisma Zh.\ Eksp.\ Teor.\ Fiz.\  {\bf 87}, 215 (2008)
  [JETP Lett.\  {\bf 87}, 185 (2008)]
  [arXiv:0803.0612 [astro-ph]].
  %%CITATION = JTPLA,87,185;%%
\bibitem{Lipovetsky}
V.~A.~Lipovetsky, S.~I.~Neizvestny and O.~M.~Neizvestnaya,
%{\it A Catalogue of Seyfert Galaxies,}
{\it Comm.\ SAO}  {\bf 55}  5 (1988)
\bibitem{Uryson}
A.~V.~Uryson,
%{\it Identification Of Extragalactic Cosmic-Ray Sources Using Data From
%Various
%Detection Facilities,} 1999
{ J.\ Exp.\ Theor.\ Phys.}\  {\bf 89} 597 (1999)
%%CITATION = JTPHE,89,597;%%
\bibitem{Uryson1}
A.~V.~Uryson,
%   ``Nearby Seyfert galaxies are possible sources of cosmic rays above
  %4*10**19-eV: Updated results,''
  arXiv:astro-ph/0303347.
 %%CITATION = ASTRO-PH/0303347;%%
\bibitem{TT:BL}
 P.~G.~Tinyakov and I.~I.~Tkachev,
  %``BL Lacertae are sources of the observed ultra-high energy cosmic rays,''
  JETP Lett.\  {\bf 74}, 445 (2001)
  [Pisma Zh.\ Eksp.\ Teor.\ Fiz.\  {\bf 74}, 499 (2001)]
  [arXiv:astro-ph/0102476].
%%CITATION = ZFPRA,74,499;%%
\bibitem{GTTT:HiRes-BL}
 D.~S.~Gorbunov, P.~G.~Tinyakov, I.~I.~Tkachev and S.~V.~Troitsky,
%   ``Testing the correlations between ultra-high-energy cosmic rays and BL
%   Lac
   %type objects with HiRes stereoscopic data,''
  JETP Lett.\  {\bf 80}, 145 (2004)
  [Pisma Zh.\ Eksp.\ Teor.\ Fiz.\  {\bf 80}, 167 (2004)]
  [arXiv:astro-ph/0406654].
%%CITATION = ASTRO-PH 0406654;%%
\bibitem{NED}
The NASA/IPAC Extragalactic database (available at
http://nedwww.ipac.caltech.edu).
\bibitem{FF}
F.~Ferrarese and H.~Ford,
%{\it Supermassive Black Holes in Galactic Nuclei: Past, Present and Future
%  Research,} 2005
{ Space Science Reviews} {\bf 116} 523 (2005)
\bibitem{0712.1630}
J.~Shen  {\it et al.},
%, D.~E.~V.~Berk, D.~P.~Schneider and P.~B.~Hall,
%  {\it The Black Hole-Bulge Relationship in Luminous Broad-Line Active
%Galactic Nuclei and Host Galaxies,} 2008
  {Astron.\ J.}\  {\bf 135} 928 (2008)
  [arXiv:0712.1630 [astro-ph]]
  %%CITATION = ANJOA,135,928;%%
\bibitem{LEDA}
G.~Paturel   {\it et al.},
%; Petit, C.; Prugniel, Ph.; Theureau, G.; Rousseau, J.; Brouty, M.;
%Dubois, P.; Cambre'sy, L.
%{\it HYPERLEDA. I. Identification and designation of
%galaxies}, 2003
{ Astron.\ Astrophys.}
  {\bf 412} 45 (2003);
http://leda.univ-lyon1.fr
\bibitem{0510694}
X.~Dong and M.~M.~De Robertis,
  %``Low-luminosity Active Galaxies and their Central Black Holes,''
  Astron.\ J.\  {\bf 131}, 1236 (2006)
  [arXiv:astro-ph/0510694].
  %%CITATION = ANJOA,131,1236;%%
\bibitem{GalacticExtinction}
D.~J.~Schlegel, D.~P.~Finkbeiner and M.~Davis,
%   ``Maps of Dust IR Emission for Use in Estimation of Reddening and CMBR
  %Foregrounds,''
  Astrophys.\ J.\  {\bf 500}, 525 (1998)
  [arXiv:astro-ph/9710327];
  %%CITATION = ASJOA,500,525;%%
http://astro.berkeley.edu/dust/index.html
\bibitem{Kband}
 Z.~X.~Peng, Q.~S.~Gu, J.~Melnick and Y.~H.~Zhao,
%, Q.~S.~Gu, J.~Melnick and Y.~H.~Zhao,
% {\it The K-band properties of Seyfert 2 galaxies,} 2006
 {Astron.\ Astrophys.} {\bf 453}
863 (2006) [arXiv:astro-ph/0603849]
%%CITATION = ASTRO-PH/0603849;%%
\bibitem{ApJ_481_L23}
  L.~J.~Greenhill, J.~M.~Moran and J.~R.~Herrnstein,
%{\it The Distribution of H$_2$O Maser Emission in the Nucleus of NGC
%  4945,} 1997
  {Astrophys.\ J.}\  {\bf 481} L23 (1997)
\bibitem{0709.3960}
R.~C.~Y.~Chou  {\it et al.},
% {\it The Circumnuclear Molecular Gas in the Seyfert Galaxy NGC4945,}
%2007
{Astrophys.\ J.}\  {\bf 670} 116 (2007)
  [arXiv:0709.3960 [astro-ph]]
\bibitem{0709.1877}
N.~Neumayer  {\it et al.},
%, M.~Cappellari, J.~Reunanen,
%H.~W.~Rix, P.~P.~van der Werf, P.~T.~de Zeeuw and R.~I.~Davies,
%{\it The central parsecs of Centaurus A: High Excitation Gas, a Molecular
%Disk,
%and the Mass of the Black Hole,}
%  2007
{Astrophys.\ J.}\  {\bf 671} 1329 (2007)
  [arXiv:0709.1877 [astro-ph]]
  %%CITATION = ARXIV:0709.1877;%%
\bibitem{ApJ_549_915}
A.~Marconi    {\it et al.},
%, Alessandro; Capetti, Alessandro; Axon, David J.; Koekemoer,
%Anton; Macchetto, Duccio; Schreier, Ethan J.
%{\it Peering through the Dust:
%        Evidence for a Supermassive Black Hole at the Nucleus of Centaurus
%        A from VLT Infrared Spectroscopy,} 2001
{Astrophys.\ J.}\ {\bf 549} 915 (2001)
\bibitem{9711137}
E.~J.~M.~Colbert  {\it et al.},
%, S.~A.~Baum, C.~P.~O'Dea and S.~Veilleux,
%{\it Large-Scale Outflows in Edge-on Seyfert Galaxies. III.
%Kiloparsec-Scale
%Soft X-ray Emission,} 1998
{Astrophys.\ J.}\ {\bf 496} 786 (1998)
[arXiv:astro-ph/9711137]
  %%CITATION = ASTRO-PH/9711137;%%
\bibitem{ApJ-537-152}
A.~L.~Kinney   {\it et al.},
%; Schmitt, H. R.; Clarke, C. J.; Pringle,
%J. E.; Ulvestad, J. S.; Antonucci, R. R. J.
%        {\it Jet Directions in Seyfert Galaxies,}
%  2000
{Astrophys.\ J.}\ {\bf 537} 152 (2000)
\bibitem{9901236}
 N.~M.~Nagar   {\it et al.},
%, A.~S.~Wilson, J.~S.~Mulchaey and J.~F.~Gallimore,
%{\it Radio Structures of Seyfert Galaxies. VIII. A Distance and Magnitude
%  Limited Sample of Early-Type Galaxies,}
%1999
{Astrophys.\ J.\ Suppl.}\ {\bf 120} 209 (1999)
[arXiv:astro-ph/9901236]
  %%CITATION = ASTRO-PH/9901236;%%
\bibitem{AA-216-39}
J.~I.~Harnett    {\it et al.},
%; Wielebinski, R.; Haynes, R. F.; Klein, U.
%        {\it Polarized radio emission from NGC 4945,}
%  1989
{Astron.\ Astrophys.}\ {\bf 216} 39 (1989)
\bibitem{0204361}
 N.~J.~Schurch, T.~P.~Roberts and R.~S.~Warwick,
%   ``High-resolution X-ray imaging and spectroscopy of the core of NGC
%  4945 with
  %XMM-Newton and Chandra,''
  Mon.\ Not.\ Roy.\ Astron.\ Soc.\  {\bf 335}, 241 (2002)
  [arXiv:astro-ph/0204361].
  %%CITATION = MNRAA,335,241;%%
\bibitem{0001459}
A.~Thean    {\it et al.},
%, A.~Pedlar, M.~J.~Kukula, S.~A.~Baum and C.~P.~O'Dea,
%{\it High-resolution radio observations of Seyfert galaxies in the
%  extended 12-micron sample - I. The observations,} 2000
  {Mon.\ Not.\
  Roy.\ Astron.\ Soc.}\  {\bf 314} 573 (2000),
  [arXiv:astro-ph/0001459]
  %%CITATION = ASTRO-PH/0001459;%%
\bibitem{AA-417-925}
  E.~Middelberg {\it et al.},
 %  ``Motion and Properties of Nuclear Radio Components in Seyfert Galaxies
%Seen
%with VLBI,''
  Astron.\ Astrophys.\  {\bf 417}, 925 (2004)
  [arXiv:astro-ph/0402142].
\bibitem{Cavallo}
G.~Cavallo,
%  On the sources of ultra-high energy cosmic rays
%1978
{\it Astron.\ Astrophys.}\ {\bf 65} 415 (1978)
\bibitem{Ancho:Cen}
  G.~E.~Romero, J.~A.~Combi, L.~A.~Anchordoqui and S.~E.~Perez Bergliaffa,
%   ``Centaurus A as a source of extragalactic cosmic rays with arrival
%  energies
  %well beyond the GZK cutoff,''
  Astropart.\ Phys.\  {\bf 5}, 279 (1996)
  [arXiv:gr-qc/9511031].
  %%CITATION = APHYE,5,279;%%
\bibitem{Farrar}
  G.~R.~Farrar and T.~Piran,
  %``Deducing the source of ultrahigh energy cosmic rays,''
  arXiv:astro-ph/0010370.
  %%CITATION = ASTRO-PH/0010370;%%
\bibitem{Weiler}
  L.~A.~Anchordoqui, H.~Goldberg and T.~J.~Weiler,
  %``An Auger test of the Cen A model of highest energy cosmic rays,''
  Phys.\ Rev.\ Lett.\  {\bf 87}, 081101 (2001)
  [arXiv:astro-ph/0103043].
  %%CITATION = PRLTA,87,081101;%%
\bibitem{AA-rev-8-237}
F.~P.~Israel,
        %{\it Centaurus A - NGC 5128,}
       %1998
{Astron.\ Astrophys.\ Rev.}\ {\bf 8} 237 (1998)
\bibitem{ChinJAAp-6-106}
        H.~Steinle,
        %{\it Centaurus A: A Multifrequency Review,}
        %2006
{Chin.\ J.\ Astron.\ Astrophys.}\ {\bf 6} 106 (2006)
\bibitem{Steinler-webpage} H.~Steinle, http://www.mpe.mpg.de/Cen-A
\bibitem{0707.0177}
  K.~Meisenheimer {\it et al.},
  %{\it Resolving the innermost parsec of Centaurus A at mid-infrared
  %wavelengths,} 2007
{Astron.\ Astrophys.}\ {\bf 471} 453 (2007)
  [arXiv:0707.0177 [astro-ph]]
  %%CITATION = ARXIV:0707.0177;%%
\bibitem{MNRAS-256-1}
        R.~Morganti    {\it et al.},
        %.; Fosbury, R. A. E.; Hook, R. N.; Robinson, A.; Tsvetanov, Z.
% {\it Evidence for a BL Lac nucleus in Centaurus A,}     1992
  {Mon.\ Not.\ Roy.\ Astron.\ Soc.}\  {\bf 256} 1 (1992)
\bibitem{0105159}
  M.~Chiaberge, A.~Capetti and A.~Celotti,
  %``The BL Lac heart of Centaurus A,''
  Mon.\ Not.\ Roy.\ Astron.\ Soc.\  {\bf 324}, L33 (2001)
  [arXiv:astro-ph/0105159].
%%CITATION = MNRAA,324,L33;%%
\bibitem{ApJ-549-L173}
J.~M.~Bai and M.~G.~Lee,
  %``Are Centaurus A And M87 Tev Gamma-Ray Sources?,''
  Astrophys.\ J.\  {\bf 549}, L173 (2001)
  [arXiv:astro-ph/0102314].
\bibitem{PASJ-52-102} PASJ-52-1021
K.~Fujisawa {\it et al.},
        %, Kenta; Inoue, Makoto; Kobayashi, Hideyuki; Murata,
        %Yasuhiro; Wajima, Kiyoaki; Kameno, Seiji; Iguchi, Satoru;
        %Horiuchi, Shinji; Sawada-Satoh, Satoko; Kawaguchi, Noriyuki; and 3
        %coauthors
        %{\it Large Angle Bending of the Light-Month Jet in Centaurus
        %A,} 2000
  {Pub.\ Astron.\ Soc.\ Japan}  {\bf 52} 1021 (2000)
\bibitem{AJ-115-960}
  S.~J.~Tingay {\it et al.},
  %.; Jauncey, D. L.; Reynolds, J. E.; Tzioumis, A. K.; King, E. A.;
   %     Preston, R. A.; Jones, D. L.; Murphy, D. W.; Meier, D. L.; van
   %     Ommen, T. D.; and 16 coauthors
%        {\it The Subparsec-Scale Structure and
%        Evolution of Centaurus A: The Nearest Active Radio Galaxy,} 1998
  {Astron.\ J.}\  {\bf 115} 960 (1998)
\bibitem{AJ-122-1697}
  S.J.~Tingay,  R.~A.~Preston and D.~L.~Jauncey,
%        {\it The Subparsec-Scale Structure and Evolution of Centaurus A.
%  II. Continued Very Long Baseline Array Monitoring,} 2001
  {Astron.\ J.}\  {\bf 115} 960 (2001)
\bibitem{ApJ-593-169}
 M.~J.~Hardcastle  {\it et al.},
  %; Worrall, D. M.; Kraft, R. P.; Forman, W. R.; Jones, C.; Murray, S. S.
%        {\it Radio and X-Ray Observations of the Jet in Centaurus A,}
%        2003
  {Astrophys.\ J.}\  {\bf 593} 169 (2003)
\bibitem{ApJ-569-54}
R.~P.~Kraft {\it et al.},
  %.; Forman, W. R.; Jones, C.; Murray, S. S.; Hardcastle, M. J.; Worrall,
  %D. M.
%  {\it Chandra Observations of the X-Ray Jet in Centaurus A,}  2002 {\it
  {Astrophys.\ J.}\  {\bf 569} 54 (2002)
\bibitem{0710.1277}
  M.~J.~Hardcastle {\it et al.},
  %{\it New results on particle acceleration in the Centaurus A jet and
  %counterjet from a deep Chandra observation,}  2007
  {Astrophys.\ J.}\  {\bf 670} L81 (2007)
  [arXiv:0710.1277 [astro-ph]]
  %%CITATION = ARXIV:0710.1277;%%
\bibitem{0510661}
  J.~Kataoka {\it et al.},
  %, L.~Stawarz, F.~Aharonian, F.~Takahara, M.~Ostrowski and
  %P.~G.~Edwards,
  %{\it The X-ray Jet in Centaurus A: Clues on the Jet Structure and
  %Particle Acceleration,}
  %2006
  {Astrophys.\ J.}\  {\bf 641} 158 (2006)
  [arXiv:astro-ph/0510661]
  %%CITATION = ASJOA,641,158;%%
\bibitem{0601421}
  M.~J.~Hardcastle, R.~P.~Kraft and D.~M.~Worrall,
%   ``The infrared jet in Centaurus A: multiwavelength constraints on
%emission
%mechanisms and particle acceleration,''
  Mon.\ Not.\ Roy.\ Astron.\ Soc.\ Lett.\  {\bf 368}, L15 (2006)
  [arXiv:astro-ph/0601421].
%%CITATION = 00482,368,L15;%%
\bibitem{ApJ-238-539}
  J.~H.~Beall and W.~K.~Rose,
%{\it On the physical environment in the nucleus of Centaurus A (NGC
%  5128),} 1980
  {\it Astrophys.\ J.}\  {\bf 238} 539 (1980)
\bibitem{0710.2847}
  J.~P.~Lenain {\it et al.},
  %, C.~Boisson, H.~Sol and K.~Katarzynski,
  %{\it A synchrotron self-Compton scenario for the very high energy
  %gamma-ray emission of the radiogalaxy M87,} 2008
  {Astron.\ Astrophys.}\  {\bf 478} 111 (2008)
  [arXiv:0710.2847 [astro-ph]]
  %%CITATION = ARXIV:0710.2847;%%
\bibitem{ApJ-395-444}
  D.~A.~Clarke, J.~O.~Burns and M.~L.~Norman,
        %{\it VLA observations of the inner lobes of Centaurus A,}
        %    1992
  {Astrophys.\ J.}\  {\bf 395} 444 (1992)
\bibitem{ApJ-273-128}
  J.~O.~Burns, E.~D.~Feigelson and E.~J.~Schreier,
        %{\it The inner radio structure of Centaurus A - Clues to the
 % origin of the jet X-ray emission,} 1983
  {Astrophys.\ J.}\  {\bf 273}
  128 (1983)
\bibitem{0709.2210}
  J.~Mao and J.~Wang,
  %{\it Knot in Cen A: Stochastic Magnetic Field for Diffusive Synchrotron
  %Radiation?,} 2007
  {Astrophys.\ J.}\  {\bf 669} L13 (2007)
  [arXiv:0709.2210 [astro-ph]]
  %%CITATION = ARXIV:0709.2210;%%
\bibitem{AA-355-863}
  H.~Alvarez {\it et al.},
  %.; Aparici, J.; May, J.; Reich, P.
%        {\it The radio continuum spectrum of Centaurus A's large-scale
%  components,} 2000
  {Astron.\ Astrophys.}\  {\bf 355} 863 (2000)
\bibitem{new-lobes}
 M.~J.~Hardcastle, C.~C.~Cheung, I.~J.~Feain and L.~Stawarz,
  %``High-energy Particle Acceleration and Production of Ultra-high-energy
  %Cosmic Rays in the Giant Lobes of Centaurus A,''
Mon.\ Not.\ Roy.\ Astron.\ Soc.\ {\bf 393} 1041 (2009)
[arXiv:0808.1593 [astro-ph]]
%%CITATION = ARXIV:0808.1593;%%
\bibitem{ApJ-646-L41}
  M.~H.~Brookes  {\it et al.},
  %.; Lawrence, C. R.; Keene, J.; Stern, D.; Gorijan, V.; Werner, M.;
  %Charmandaris, V.
  %{\it Spitzer Observations of Centaurus A: Infrared
  %      Synchrotron Emission from the Northern Lobe,} 2006
  {Astrophys.\ J.}\  {\bf 646} L41 (2006)
\bibitem{new-PAO-talk}
D.~Martello et al.\ [Pierre Auger Collaboration], Talk at Scineghe 2009,
Assisi, Italy; available from
{\tt
http://agenda.infn.it/materialDisplay.py?contribId=27\&{}amp;
sessionId=8\&{}amp;materialId=slides\&{}amp;confId=1369}

\end{thebibliography}
\end{document}